\documentclass[graybox]{svmult}

\usepackage{type1cm}        
\usepackage{makeidx}         
\usepackage{graphicx}        
\usepackage{multicol}        
\usepackage[bottom]{footmisc}

\usepackage[normalem]{ulem}
\usepackage{newtxtext}       %
\usepackage{newtxmath}       
\usepackage{float}

\usepackage{subfig}
\usepackage{epsf}
\usepackage{epsfig}
\usepackage{epstopdf}
\usepackage{wrapfig}
\usepackage{float}
\usepackage{threeparttable}
\usepackage{makecell}
\usepackage{mathtools}
\usepackage[scientific-notation=true]{siunitx}

\makeindex             

\begin{document}

\title*{Linear Instability of Shock-Dominated Laminar Hypersonic Separated Flows}
\author{Saurabh S. Sawant, Ozgur Tumuklu, Vassilis Theofilis, and Deborah A. Levin}
\institute{Deborah A. Levin \at University of Illinois at Urbana-Champaign, USA, \email{deblevin@illinois.edu}}
\institute{Vassilis Theofilis \at University of Liverpool, UK, \email{v.theofilis@liverpool.ac.uk}}
\institute{Ozgur Tumuklu \at University of Illinois at Urbana-Champaign, USA, \email{tumuklu2@illinois.edu}}
\institute{Saurabh S. Sawant \at University of Illinois at Urbana-Champaign, USA, \email{sssawan2@illinois.edu}}
%
%
\maketitle

\abstract*{Each chapter should be preceded by an abstract (no more than 200 words) that summarizes the content. The abstract will appear \textit{online} at \url{www.SpringerLink.com} and be available with unrestricted access. This allows unregistered users to read the abstract as a teaser for the complete chapter.
Please use the 'starred' version of the \texttt{abstract} command for typesetting the text of the online abstracts (cf. source file of this chapter template \texttt{abstract}) and include them with the source files of your manuscript. Use the plain \texttt{abstract} command if the abstract is also to appear in the printed version of the book.}

\abstract{The self-excited spanwise homogeneous perturbations arising in shock-wave/boundary-layer interaction (SWBLI) system formed in a hypersonic flow of molecular nitrogen over a double wedge are investigated using the kinetic Direct Simulation Monte Carlo (DSMC) method.
The flow has transitional Knudsen and unit Reynolds numbers of~\num{3.4e-3} and \num{5.2e4}~m$^{-1}$, respectively.
Strong thermal nonequilibrium exists downstream of the Mach 7 detached (bow) shock generated due to the upper wedge surface. 
A linear instability mechanism is expected to make the pre-computed 2-D base flow potentially unstable under spanwise perturbations.
The specific intent is to assess the growth rates of unstable modes, the wavelength, location, and origin of spanwise periodic flow structures, and the characteristic frequencies present in this interaction.}

\section{Introduction}
The identification and analysis of potential growing modes in unsteady, laminar hypersonic flows is essential to predict the possible transition to turbulence, which in turn affects the aerothermodynamic loads on the surface of an embedded body.
This work concerns an Edney type-IV/V SWBLI system~\cite{edney1968anomalous} in a molecular nitrogen flow over a double wedge which exhibits complex flow dynamics characterized by thermal nonequilibrium, a flow recirculation zone near the hinge of the two angular surfaces, shear layers downstream of the triple point and above the recirculation zone, and $\lambda$-shocklets~\cite{delery_2011}.
In particular, understanding of these phenomena at unit Reynolds number, $Re_1$, on the order of 10$^5$ m$^{-1}$ using the stochastic DSMC method~\cite{bird:94mgd} provides highest fidelity for shock dominated flows.
The DSMC method is valid for a broader range of Knudsen numbers than Navier-Stokes equations; as a result, the critical treatment of strong shocks (M$\gg$1.6)~\cite{bird1970aspects,ohwada1993structure,cercignani1999structure}, the rarefaction effects such as the velocity and temperature slip~\cite{sawant2018application,tumuklu2016factors,moss2005direct,chambre2017flow}, thermal nonequilibrium~\cite{bird:94mgd,sawant2018application}, and the temporal evolution of self-excited perturbations are natural outcomes of the DSMC approach~\cite{tumuklu2018POF1, tumuklu2018POF2,tumuklu2018PhysRevF}.\vspace{\baselineskip}

\begin{wrapfigure}{r}{0.45\textwidth}
\vspace{-2em}
  \includegraphics[width=0.45\textwidth]{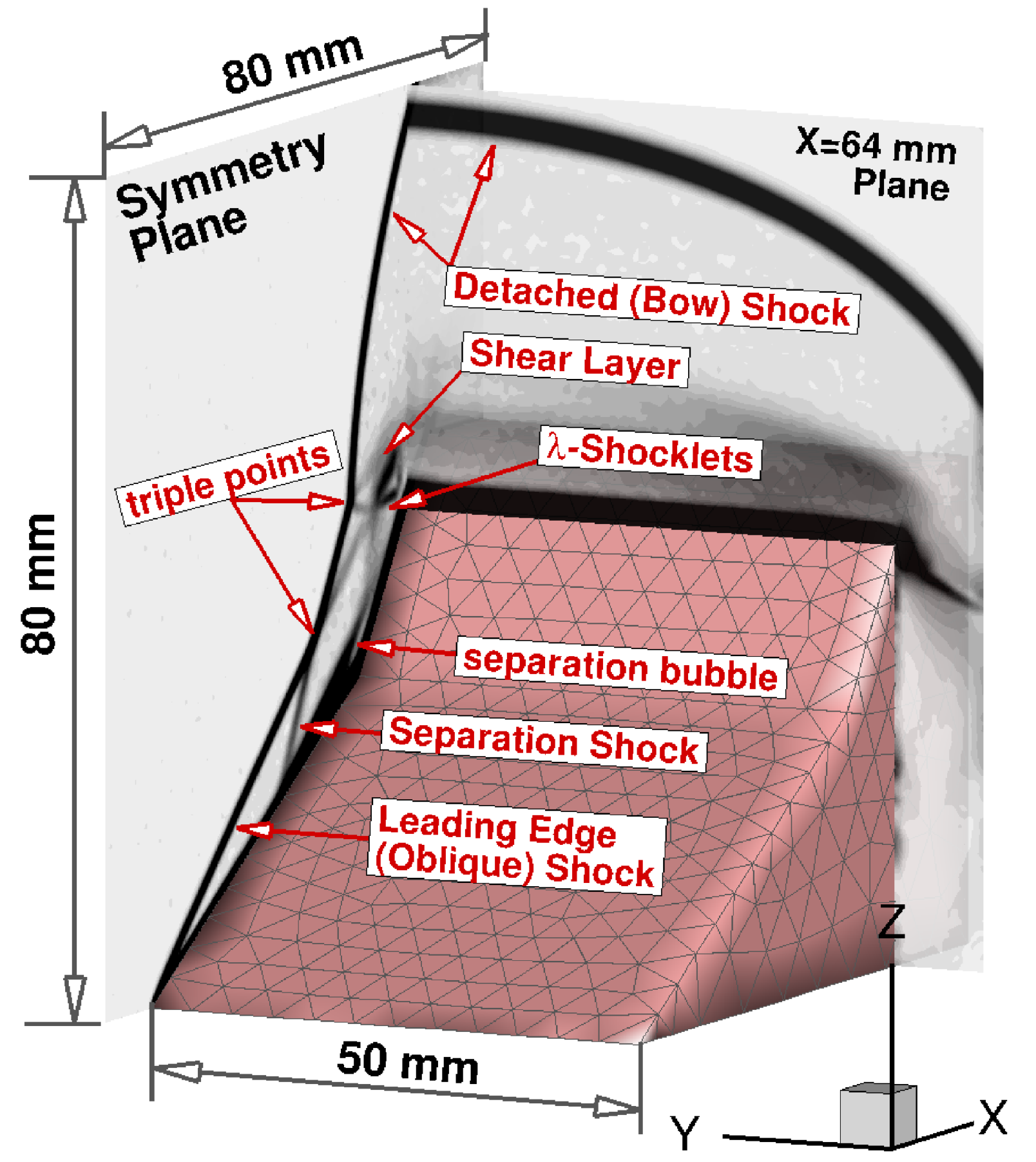}
 \caption{SWBLI interactions for the previous, three dimensional (3-D), symmetry-imposed, hypersonic flow of nitrogen over the double wedge~\cite{swantek2014} at 0.4~ms~\cite{sawant2018application}.}
 \label{f:1}
\end{wrapfigure}
\begin{figure}[H]
\begin{center}
\includegraphics[width=0.80\columnwidth]{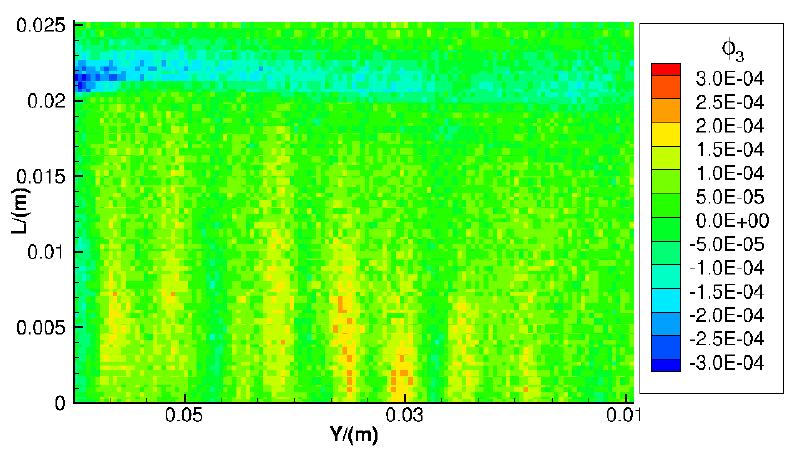}
\caption{
The contours show the third spatial mode on a slant plane 3.44~mm above and parallel to the upper wedge and between the boundary and shear layers. Along the span, the symmetry plane and spanwise opposite edge are at Y=60 and at 9.2~mm, respectively.}
\label{f:1dot5}
\end{center}
\end{figure}
\begin{wrapfigure}{r}{0.45\textwidth}
\vspace{-2em}
  \includegraphics[width=0.45\textwidth, height=0.45\textwidth]{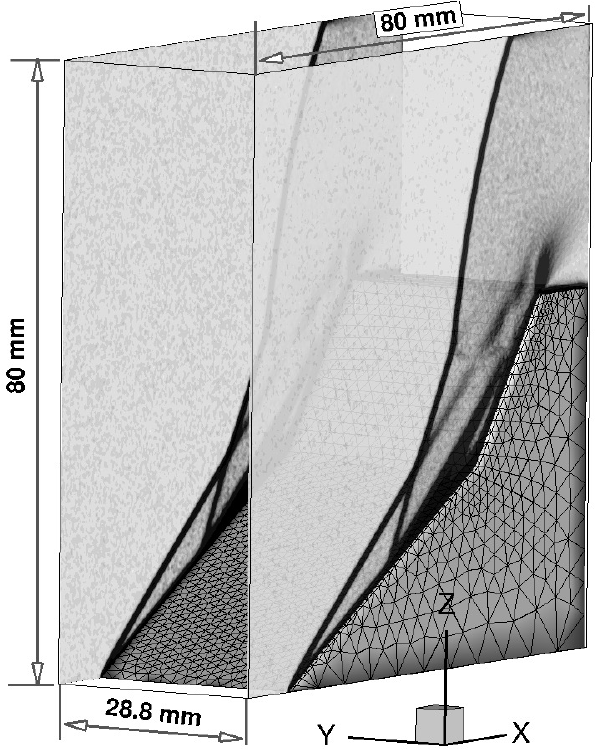}
  \caption{New, span-periodic setup.}
  \label{f:2}
\end{wrapfigure}

Tumuklu et al.~\cite{tumuklu2018PhysRevF} studied a 2-D (spanwise independent) flow over a double wedge and demonstrated decaying modes using proper orthogonal decomposition (POD).
The flow took 0.9~ms to reach steady-state and the recirculation zone of 36~mm length was developed at the hinge.
Previously, the authors studied a three dimensional (3-D) flow over a double wedge with a finite half-span of 50.8~mm~\cite{sawant2018application} by imposing a symmetry plane at the center of the span, as shown in Fig.~\ref{f:1}.
The flow was found to be unsteady for 0.57~ms of simulation and started to exhibit 3-D structures inside the separation bubble and at the reattachment point.
Figure~\ref{f:1dot5} shows the third spatial mode obtained from a POD analysis of 0.4 to 0.57~ms DSMC data of streamwise velocity. 
Spanwise striations are less prominent towards the free edge of the wedge because of the spanwise acceleration of the gas can be observed.
It is challenging to understand the origin of such striations and to determine their characteristic wavelength in the presence of flow pressure relief induced by the finite span.
Therefore, to understand the unsteady dynamics of this SWBLI system, a simpler, spanwise periodic case has been undertaken~\cite{sawant2019correction}, as shown in Fig.~\ref{f:2} describing the simulation setup.\vspace{\baselineskip}

The specific goal of this work is first to understand whether the 2-D base flow remains stable under spanwise homogeneous, self-excited perturbations.
The presence of strong recirculation zones and shear layers is a recipe for generation and/or amplification of self-excited spanwise perturbations through linear instability mechanisms~\cite{theofilis2000origins, theofilis2011global, pagella2004instability, sidharth2018onset,dwivedi2019reattachment, alves2019steady, gai2019hypersonic}.
If these perturbations do not damped out, naturally, then the amplitude of the disturbance characteristic wavelength is amplified in the SWBLI system and resulting spanwise periodic flow structures occur with a periodicity of this wavelength~\cite{theofilis2011global, sidharth2018onset, Chuvakhov2017, Roghelia2017}.
If this occurs, the spatial and temporal characteristics of such structures, their origin, and growth rate will be investigated in this work.

DSMC performs advection of simulated particles, each of which represents a large number of real molecules, for a duration of a discrete timestep~\cite{bird:94mgd}.
Then the particles are mapped on to discrete cells of the collision mesh encompassing the flow domain.
From each cell, particle pairs are selected using appropriate collision cross-sections, and post-collisional instantaneous velocities are computed.
Note that the particles also reflect from embedded surfaces using gas-surface collision models and are introduced, removed, or reflected from the domain boundaries.
The macroscopic flow parameters such as velocity and temperatures are extracted from the particles' instantaneous velocities using statistical relations of kinetic theory~\cite{bird:94mgd,vincenti1965introduction,koganRGDSpringer}.
In complex 3-D flows, billions of computational particles are simulated to satisfy the numerical requirements.
This is achieved by the Scalable Unstructured Gas-dynamics Adaptive mesh-Refinement (SUGAR-3D) code that has been demonstrated to scale well on many thousands of CPU-cores~\cite{sawant2018application}.
By using SUGAR-3D, a preliminary spanwise periodic case was simulated~\cite{sawant2019correction} for 0.25~ms with a span length of 72~mm, (twice the size of the separation length), which revealed 7.2~mm long spanwise periodic structures downstream of reattachment.
With this estimate, a 28.8~mm span case was simulated using 19,200 processors with $\sim$60~billion particles and $\sim$4.5~billions collision cells in a (400 $\times$ 144 $\times$ 400) octree grid.
The input parameters are given in table~\ref{tab:1}, whereas the convergence and span independence is demonstrated in Appendix.
\begin{table}[H]
\caption{Simulation parameters for a flow of molecular nitrogen over a double wedge.}
\label{tab:1}       
\begin{tabular}{p{8cm}p{3.5cm}}
\hline\noalign{\smallskip}
Parameters & Values   \\
\noalign{\smallskip}\svhline\noalign{\smallskip}
Mach number, M                                                    & 7 \\
Unit Reynolds number                                              & 52,200 \\
Freestream Knudsen number                                         & 0.0034 \\
Freestream number density, n/(m$^3$)                              & \num{1e22} \\
Freestream gas temperatures$^a$/(K)   & 710           \\
Surface$^b$ temperature, $T_s$/(K)                                & 298.5      \\
Timestep, $\Delta t$/(ns)                                         & 5           \\
Mesh refinement interval/($\mu$s)                                          & 5           \\
\noalign{\smallskip}\hline\noalign{\smallskip}
\end{tabular}
$^a$ Translational, rotational, and vibrational temperatures are denoted as $T_{tr}$, $T_{rot}$, and $T_{vib}$, respectively. For thermal relaxation Larsen-Borgnakke model with Millikan-White expression for vibrational probability is used. 
$^b$ The surface is fully accommodated.
\end{table}
\section{Small-Amplitude Disturbances}

\begin{figure}[H]
    \centering
    \subfloat[Iso-contours of $|KE|_2$.]{\label{f:4a}{\includegraphics[width=0.5\textwidth]{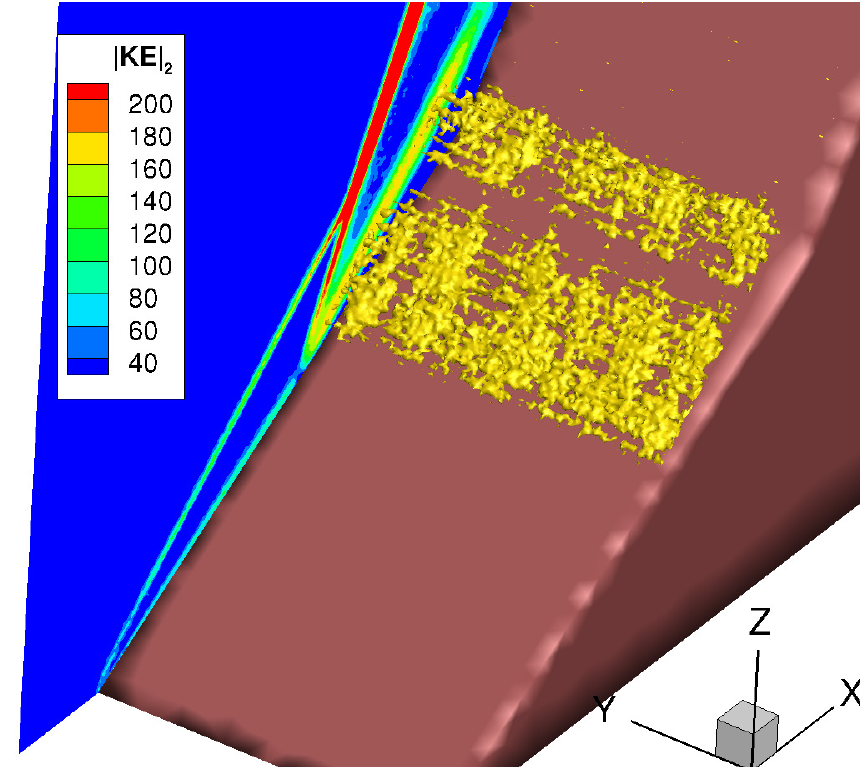}}}\hfill
    \subfloat[$|KE|_2$ at probe `S$_5$', denoted in Fig.~\ref{f:5a}.]{\label{f:4b}{\includegraphics[width=0.5\textwidth]{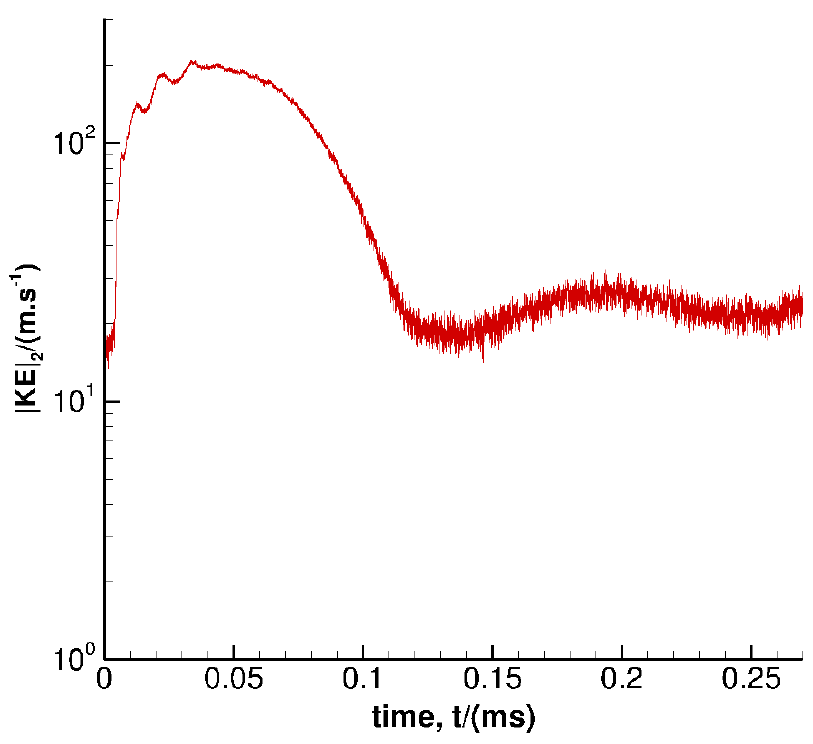}}}\hfill
    \caption{Figure~\ref{f:4a} shows spanwise periodic structures in the shear layer above the separation bubble at time t=0.06~ms and Fig.~\ref{f:4b} shows the temporal history of spanwise averaged residual kinetic energy.}
    \label{f:4}
\end{figure}

The small amplitude disturbances are seen to grow in the recirculation region and the shear layer above it.
Figure~\ref{f:4a} shows the iso-contours of residual kinetic energy defined by, 
\begin{equation} 
|KE|_2 =\sqrt{(U-U_{b})^2 + V^2 + (W-W_{b})^2}
\label{Eq:RKE}
\end{equation}
where $U$, $V$, and $W$ are bulk velocities in the streamwise ($X$), spanwise ($Y$), and streamwise-normal ($Z$) directions, respectively, whereas $U_b$ and $W_b$ are respective 2-D base flow values at a given $X-Z$ location.
Six spanwise periodic structures can be identified in the shear layer above the separation bubble.
Note that to plot the iso-contours, the velocities in Eq.~\ref{Eq:RKE} are sampled only for 200 timesteps (1$\mu$s) from the time instant of t=0.06~ms and as a result, suffer from statistical scatter.
The time instant t=0.06~ms is roughly the end of transient period, as seen from Fig.~\ref{f:4b}, which shows the temporal history of residual kinetic energy.
These structures change dynamically on the same time scale; therefore, they cannot be sampled for a longer duration.
Later, as the disturbance amplitude increases, such structures are expected to become more prominent.
Furthermore, from Fig.~\ref{f:4b}, a low frequency sinusoidal variation is observed from 0.13~ms.
Such variation is also seen at probes located at the separation point, which is analyzed in Sec.~\ref{FrequencyAnalysis}.

\section{Frequency Analysis}~\label{FrequencyAnalysis}
\begin{figure}[H]
    \centering
    \subfloat[Numerical probes overlaid on contours of streamwise velocity at Y=14.4~mm.]{\label{f:5a}{\includegraphics[width=0.45\textwidth]{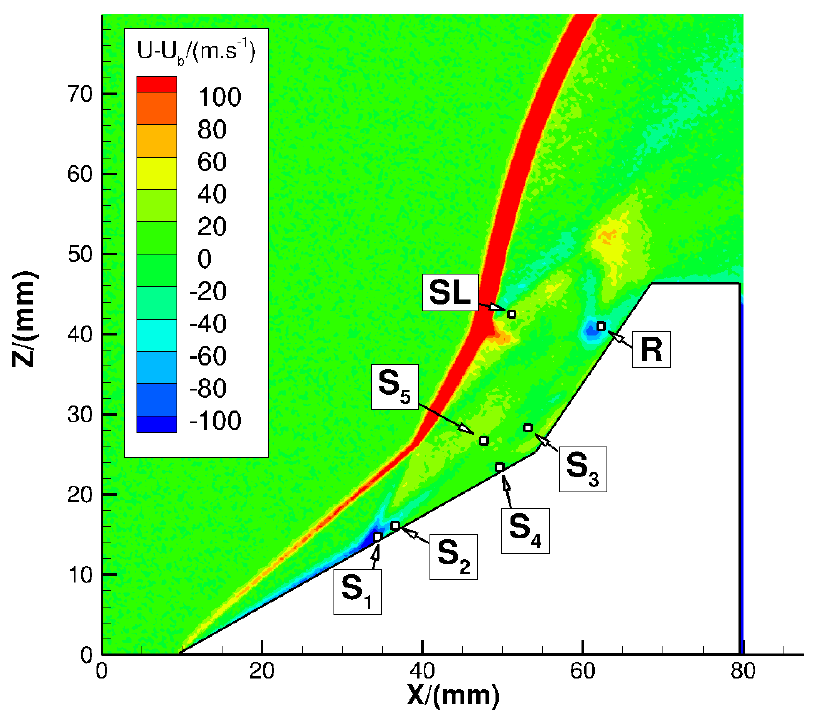}}}\hfill
    \subfloat[Pressure residual at probes `R', `SL', and `S$_3$'.]{\label{f:5b}{\includegraphics[width=0.45\textwidth]{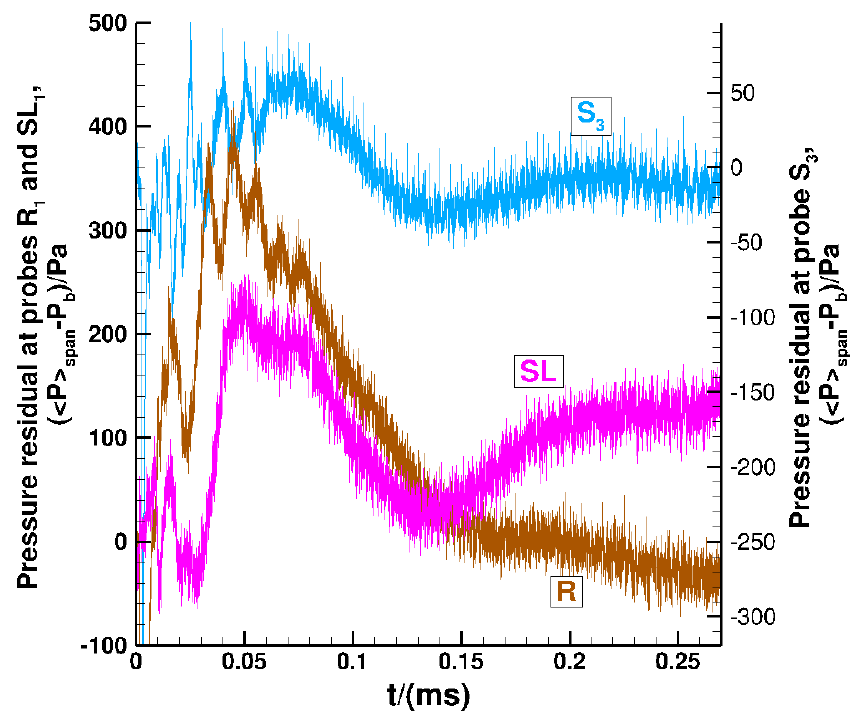}}}\,
    \subfloat[Low frequency oscillations at the separation point.]{\label{f:5c}{\includegraphics[width=0.45\textwidth]{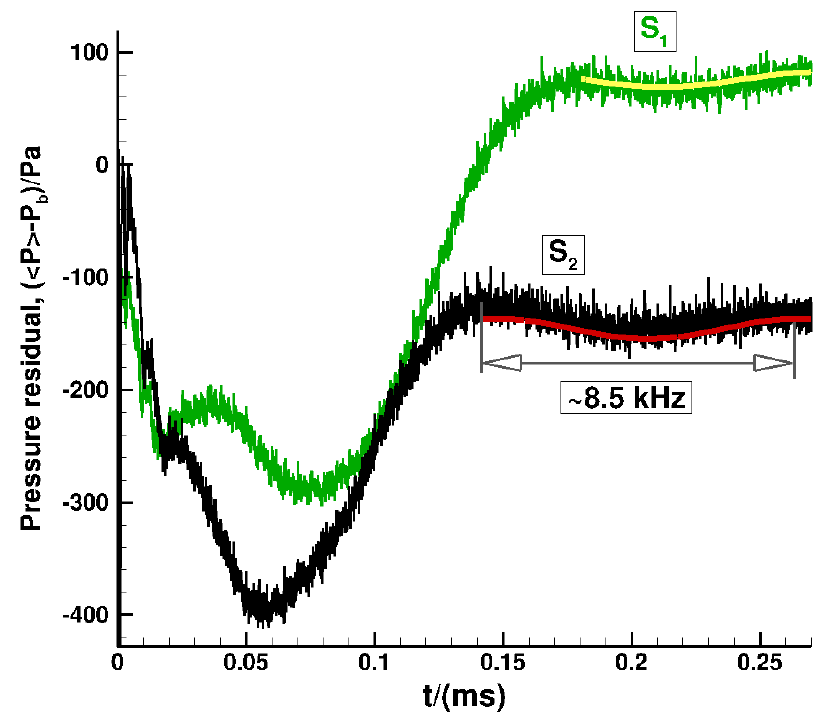}}}\hfill
    \subfloat[PSD for data at probe `R'.]{\label{f:5d}{\includegraphics[width=0.45\textwidth]{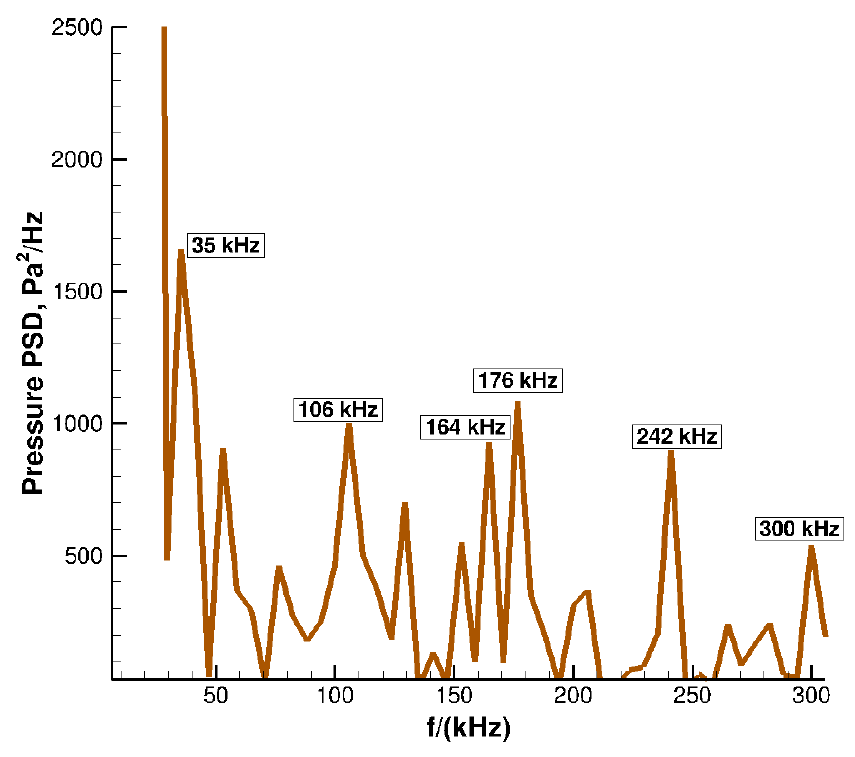}}}\hfill
    \caption{Underlying frequencies in the temporal history of spanwise averaged pressure residual at numerical probe locations denoted in Fig.~\ref{f:5a} inside and at the foot of separation bubble, at reattachment, and inside shear layer downstream of the triple point.}
    \label{f:5}
\end{figure}

Spanwise averaged pressure residual at the reattachment location, shear layer downstream of the triple point, and inside the separation region is shown in Fig.~\ref{f:5b}.
It is seen that flow is unsteady, and the case needs to be run further to understand the long-time behavior of flow evolution.
Notice a dip in the residuals at nearly 0.135~ms in the shear layer and the separation indicating that these dynamics are coupled.
Such a decrease is also seen at reattachment at 0.16~ms, indicating a delayed response to the region of reattachment.
Inside the separation region, a low-frequency wave is seen to originate after the decrease at 0.135~ms.
An 8.5~kHz of frequency oscillation is seen at probe S$_2$ at the foot of separation in Figs.~\ref{f:5c}.
Later, because the separation shock moved upstream, the delayed frequency oscillation is seen to develop in probe S$_1$.
Furthermore, to understand the high frequencies present in residuals, the power spectral density (PSD) of the pressure spectra from 0.1~ms to 0.27~ms was computed.
At the reattachment and shear layer, many high frequencies ranging from 30 to 300~kHz are observed.
Figure~\ref{f:5d} shows a dominant frequency of 35~kHz at reattachment, whereas, in the shear layer, it is 46~kHz (not shown).
These frequencies also exist inside the separation bubble (not shown). 
Note that the frequency resolution for the current analysis is 6~kHz.
Further analyses performed with a finer resolution on probe data obtained at longer times (0.9~ms) will be described in a journal paper~\cite{DWJFM_Future}.

\newpage
\section{Conclusion}
In summary, the self-excited spanwise perturbations are formed in the shear layer above the separation bubble, and its imprint is noticeable in spite of the statistical scatter present in the stochastic DSMC method.
Spanwise periodic structures appear intermittently but are prominent in the shear layer above the separation bubble.
A presence of 8.5~kHz low frequency is observed in residuals of streamwise velocity and pressure at the location of the shear layer above and inside the separation bubble.
Besides, a range of high frequencies between 30 to 300~kHz are present in the shear layer downstream of triple point, at reattachment, and inside the separation bubble.
In particular, 35$\pm$6 and 46$\pm$6~kHz dominant frequencies are observed in the PSD of pressure spectra at reattachment and shear layer downstream of the triple point.
A detailed analysis confirming the prediction of the linear nature of instability will be presented in a journal paper, shortly.
In addition, the identification of the nature of such disturbances (acoustic, vorticity waves, entropy waves)~\cite{duck1995interaction} will be studied in future.

%

\begin{acknowledgement}
This research is part of the Blue Waters sustained-petascale computing project, which is supported by the National Science Foundation (awards OCI-0725070 and ACI-1238993) and the state of Illinois. Blue Waters is a joint effort of the University of Illinois at Urbana-Champaign and its National Center for Supercomputing Applications.
\end{acknowledgement}
\newpage
\section*{Appendix}~\label{Appendix}
\begin{figure}[H]
    \centering
    \subfloat[Ratio of local mean-free-path to the collision cell size.]{\label{f:3a}{\includegraphics[width=0.5\textwidth]{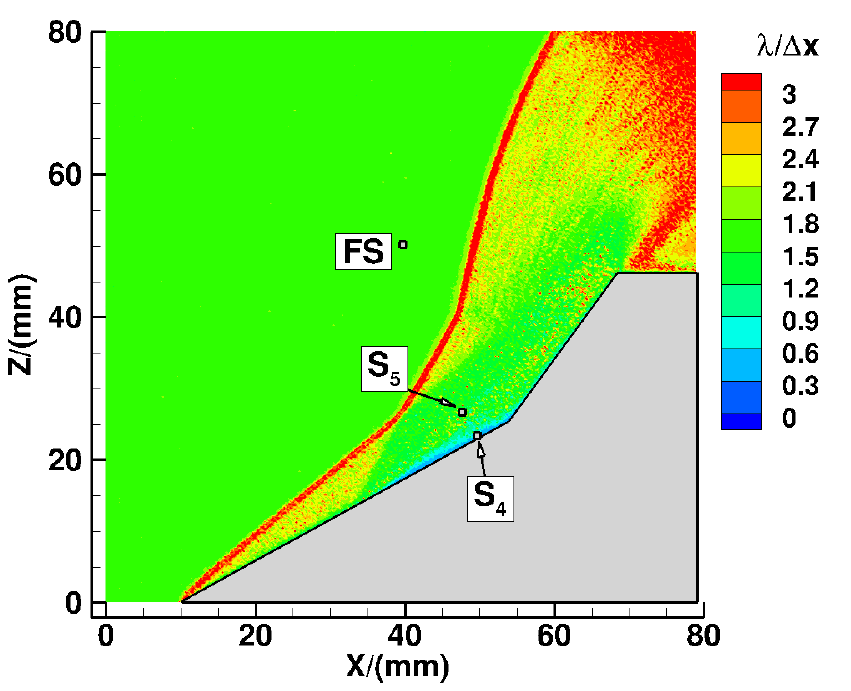}}}\hfill
    \subfloat[Ratio of local mean-collision-time to the timestep.]{\label{f:3b}{\includegraphics[width=0.5\textwidth]{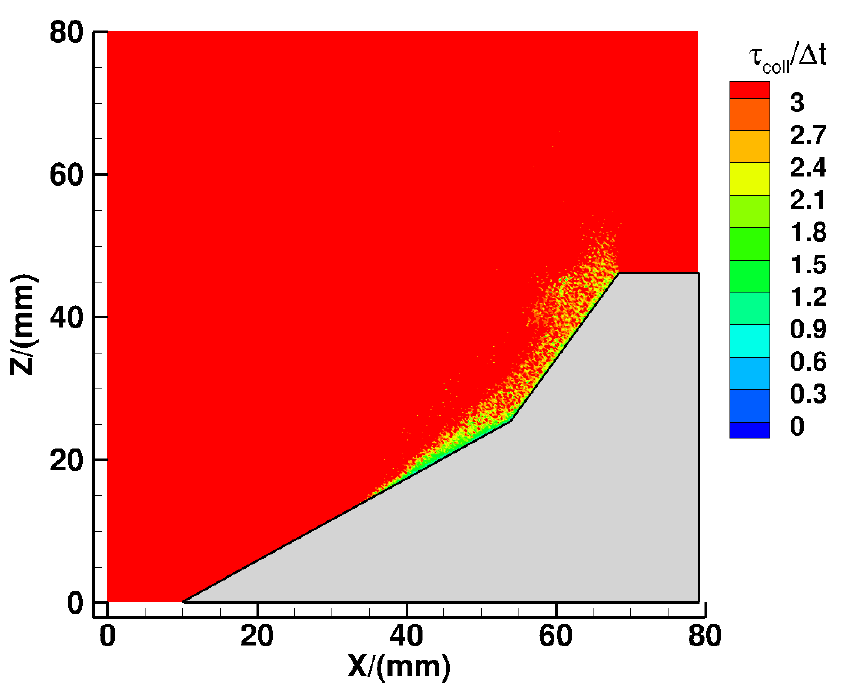}}}\,
    \subfloat[Number of computational particles per collision cell.]{\label{f:3c}{\includegraphics[width=0.5\textwidth]{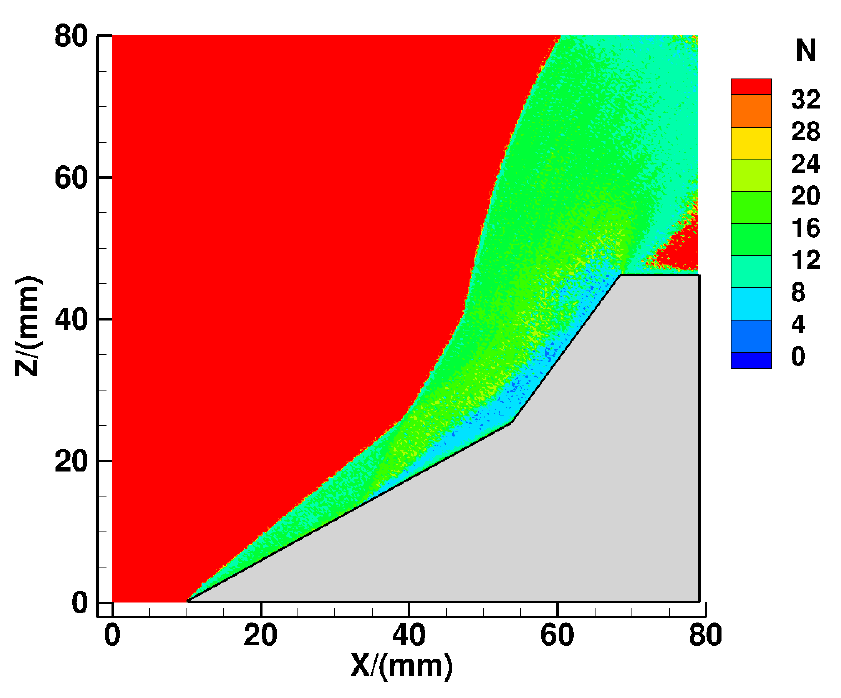}}}\hfill
    \subfloat[Spanwise independence in spanwise averaged translational temperature.]{\label{f:3e}{\includegraphics[width=0.5\textwidth]{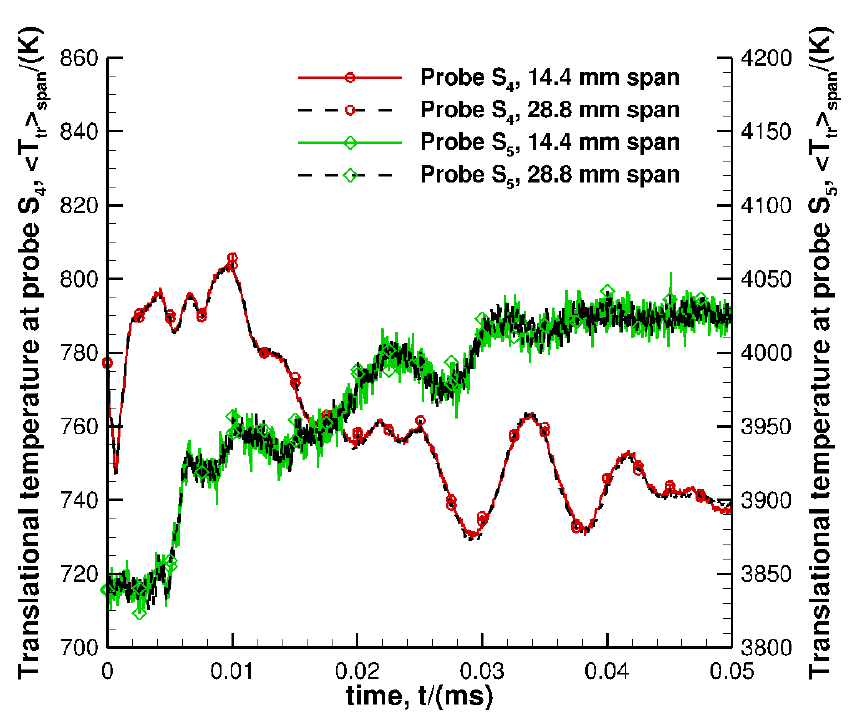}}}\,
    \subfloat[Convergence of spanwise averaged streamwise velocity (U) at locations denoted in Fig.~\ref{f:3a}.]{\label{f:3d}{\includegraphics[width=0.5\textwidth]{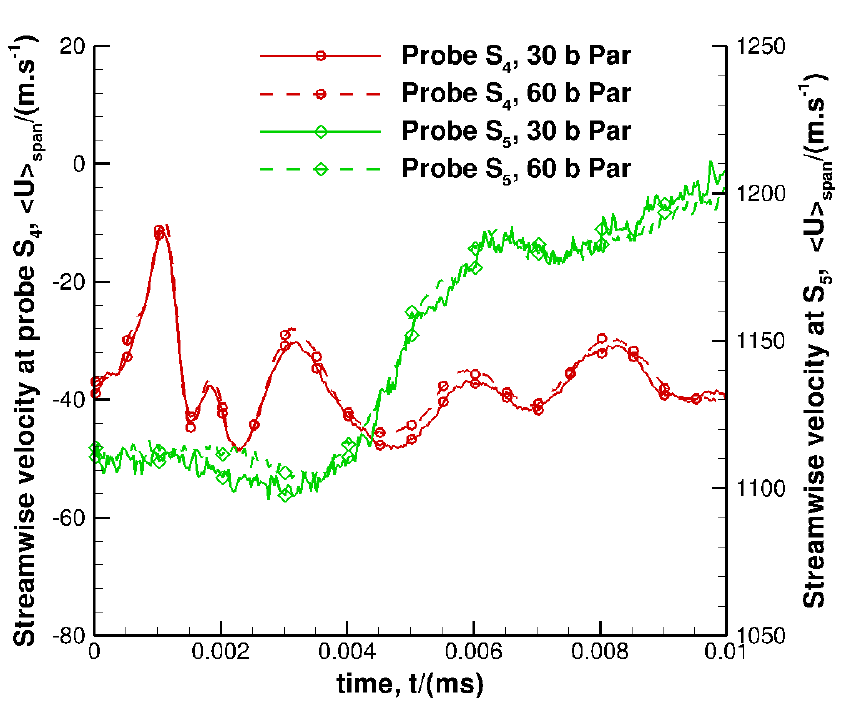}}}
    \caption{Convergence and spanwise independence of the spanwise periodic simulation. The X-Z Contours in Fig.~\ref{f:3a},~\ref{f:3b},~\ref{f:3c} are shown at Y=14.4~mm for the 28.8~mm span case with 60~billion particles. Figure~\ref{f:3e} shows spanwise independence between the 28.8~and 14.4~mm span cases with 60~and 30~billion particles, respectively. Figure~\ref{f:3d} shows convergence in terms of the number of particles for the 14.4~mm span case.}
    \label{f:3}
\end{figure}

Figure~\ref{f:3a} shows that local mean-free-path is comparable to the collision cell size in the vicinity of the wedge surface inside the recirculation region and is greater than one everywhere else, including the strong shocks.
To ensure the decoupling of collisions and movement of particles, the ratio of mean-collision-time to timestep is much higher than one, as shown in Fig.~\ref{f:3b}.
To ensure unbiased collisions, the number of computational particles per collision cell are at least eight in the recirculation region, as shown in Fig.~\ref{f:3c}, where the cell size is \num{1.25e-5} (fourth level of refinement).
To ensure that the numerical results are independent of the chosen span length of 28.8~mm (60~billion particles), another case with 14.4~mm span length (30~billion particles) was run, and the comparison is shown in Fig.~\ref{f:3d}.
The comparison of spanwise averaged translational temperature as a function of time is excellent at probe locations S$_4$ and S$_5$ in the vicinity of surface and further away in the shear layer above the separation bubble.
Furthermore, a stringent test of convergence is performed on a 14.4~mm span case (30~billion particles), where the effect of doubling the number of particles is compared in Fig.~\ref{f:3e}.
It is seen that the spanwise averaged velocity follows the same temporal trend during the initial transient phase, where severe changes in macroscopic flow parameters are seen.
Close to the surface where the Knudsen number is 0.9, the maximum difference in spanwise averaged velocity is 5 to 7\% at the peaks, whereas away from the surface, the difference is within 1.5\%.
\addcontentsline{toc}{section}{Appendix}

\bibliography{References}

\begin{thebibliography}{10}
\providecommand{\url}[1]{{#1}}
\providecommand{\urlprefix}{URL }
\expandafter\ifx\csname urlstyle\endcsname\relax
  \providecommand{\doi}[1]{DOI \discretionary{}{}{}#1}\else
  \providecommand{\doi}{DOI \discretionary{}{}{}\begingroup
  \urlstyle{rm}\Url}\fi

\bibitem{edney1968anomalous}
B.~Edney, Anomalous heat transfer and pressure distributions on blunt bodies at
  hypersonic speeds in the presence of an impinging shock.
\newblock Tech. rep., Flygtekniska Forsoksanstalten, Stockholm (Sweden) (1968)

\bibitem{delery_2011}
J.~Délery, \emph{Physical Introduction} (Cambridge University Press, 2011), p.
  5–86.
\newblock Cambridge Aerospace Series.
\newblock \doi{10.1017/CBO9780511842757.002}

\bibitem{bird:94mgd}
G.A. Bird, \emph{Molecular Gas Dynamics and the Direct Simulation of Gas
  Flows}, 2nd edn.
\newblock Oxford Engineering Science Series (Clarendon Press, 1994)

\bibitem{bird1970aspects}
G.A. Bird, The Physics of Fluids \textbf{13}(5), 1172 (1970).
\newblock \doi{10.1063/1.1693047}

\bibitem{ohwada1993structure}
T.~Ohwada, Physics of Fluids A: Fluid Dynamics \textbf{5}(1), 217 (1993).
\newblock \doi{10.1063/1.858777}

\bibitem{cercignani1999structure}
C.~Cercignani, A.~Frezzotti, P.~Grosfils, Physics of fluids \textbf{11}(9),
  2757 (1999).
\newblock \doi{10.1063/1.870134}

\bibitem{sawant2018application}
S.S. Sawant, O.~Tumuklu, R.~Jambunathan, D.A. Levin, Computers \& Fluids
  \textbf{170}, 197 (2018).
\newblock \doi{10.1016/j.compfluid.2018.04.026}

\bibitem{tumuklu2016factors}
O.~Tumuklu, D.A. Levin, S.F. Gimelshein, J.M. Austin, in \emph{AIP Conference
  Proceedings}, vol. 1786 (AIP Publishing, 2016), vol. 1786, p. 050005.
\newblock \doi{10.1063/1.4967555}

\bibitem{moss2005direct}
J.N. Moss, G.A. Bird, AIAA journal \textbf{43}(12), 2565 (2005).
\newblock \doi{10.2514/1.12532}

\bibitem{chambre2017flow}
P.A. Chambre, S.A. Schaaf, \emph{Flow of rarefied gases}, vol. 4971 (Princeton
  University Press, 2017)

\bibitem{tumuklu2018POF1}
O.~Tumuklu, D.A. Levin, V.~Theofilis, Physics of Fluids \textbf{30}(4), 046103
  (2018).
\newblock \doi{10.1063/1.5022598}

\bibitem{tumuklu2018POF2}
O.~Tumuklu, V.~Theofilis, D.A. Levin, Physics of Fluids \textbf{30}(10), 106111
  (2018).
\newblock \doi{10.1063/1.5047791}

\bibitem{tumuklu2018PhysRevF}
O.~Tumuklu, D.A. Levin, V.~Theofilis, Physical Review Fluids \textbf{4}(3),
  033403 (2019).
\newblock \doi{10.1103/PhysRevFluids.4.033403}

\bibitem{swantek2014}
A.~Swantek, J.~Austin, AIAA Journal \textbf{53}(2), 311 (2014)

\bibitem{sawant2019correction}
S.S. Sawant, O.~Tumuklu, V.~Theofilis, D.A. Levin, in \emph{AIAA Aviation 2019
  Forum} (2019), pp. 3442--c1.
\newblock \doi{10.2514/6.2019-3442.c1}

\bibitem{theofilis2000origins}
V.~Theofilis, S.~Hein, U.~Dallmann, Philosophical Transactions of the Royal
  Society of London. Series A: Mathematical, Physical and Engineering Sciences
  \textbf{358}(1777), 3229 (2000).
\newblock \doi{10.1098/rsta.2000.0706}

\bibitem{theofilis2011global}
V.~Theofilis, Annual Review of Fluid Mechanics \textbf{43}, 319 (2011).
\newblock \doi{10.1146/annurev-fluid-122109-160705}

\bibitem{pagella2004instability}
A.~Pagella, U.~Rist, in \emph{Meeting Proceedings RTO-MP-AVT-111, Prague, Czech
  Republic} (2004).
\newblock \doi{10.1007/BF00311809}

\bibitem{sidharth2018onset}
G.~Sidharth, A.~Dwivedi, G.V. Candler, J.W. Nichols, Physical Review Fluids
  \textbf{3}(9), 093901 (2018).
\newblock \doi{10.1103/PhysRevFluids.3.093901}

\bibitem{dwivedi2019reattachment}
A.~Dwivedi, G.~Sidharth, J.W. Nichols, G.V. Candler, M.R. Jovanovi{\'c},
  Journal of Fluid Mechanics \textbf{880}, 113 (2019).
\newblock \doi{10.1017/jfm.2019.702}

\bibitem{alves2019steady}
L.S. Alves, R.D. Santos, N.~Cerulus, V.~Theofilis, in \emph{AIAA Scitech 2019
  Forum} (2019), p. 2321.
\newblock \doi{10.2514/6.2019-2321}

\bibitem{gai2019hypersonic}
S.L. Gai, A.~Khraibut, Journal of Fluid Mechanics \textbf{877}, 471 (2019).
\newblock \doi{10.1017/jfm.2019.599}

\bibitem{Chuvakhov2017}
P.V. Chuvakhov, V.Y. Borovoy, I.V. Egorov, V.N. Radchenko, H.~Olivier,
  A.~Roghelia, Journal of Applied Mechanics and Technical Physics
  \textbf{58}(6), 975 (2017).
\newblock \doi{10.1134/S0021894417060037}

\bibitem{Roghelia2017}
A.~Roghelia, P.V. Chuvakhov, H.~Olivier, I.V. Egorov, 47th AIAA Fluid Dynamics
  Conference (June), 1 (2017).
\newblock \doi{10.2514/6.2017-3463}

\bibitem{vincenti1965introduction}
W.G. Vincenti, C.H. Kruger, \emph{Introduction to physical gas dynamics}
  ({Wiley, New York}, 1965)

\bibitem{koganRGDSpringer}
M.~Kogan, \emph{Rarefied Gas Dynamics}, 1st edn. (Springer, 1969)

\bibitem{DWJFM_Future}
S.S. Sawant, V.~Theofilis, D.A. Levin, Journal of Fluid Mechanics (in
  preparation)  (2020)

\bibitem{duck1995interaction}
P.W. Duck, D.G. Lasseigne, M.~Hussaini, Theoretical and Computational Fluid
  Dynamics \textbf{7}(2), 119 (1995).
\newblock \doi{10.1007/BF00311809}

\end{thebibliography}
\bibliographystyle{spphys}

\end{document}